\documentstyle[12pt]{article}
\begin{document}
\rightline{YCTP-P21-92}
\baselineskip=15pt
\rightline{June 1992}
\rightline{revised July 1992}
\baselineskip=21pt

\vskip .2in
\begin{center}
{\large{\bf COBE, INFLATION, AND LIGHT SCALARS}}
\end{center}
\vskip .1in
\begin{center}
Lawrence M.
Krauss\footnote{Bitnet: Krauss@Yalehep. Research supported in part
by the NSF,DOE, and TNRLC}

{\it Center for Theoretical Physics and
Dept of Astronomy}

{\it Sloane Laboratory}

{\it  Yale University, New Haven, CT 06511}

\end{center}

\vskip .2in

\centerline{ {\bf Abstract} }

\baselineskip=18pt

\noindent
Comparison of the COBE observed quadrupole anisotropy
with that predicted in adiabatic CDM models suggests
that much of the observed signal may be due to
long wavelength gravitational waves.  In inflationary models this
requires
the generation of tensor fluctuations to be at least
comparable to scalar density fluctuations.  This is feasible, but
depends sensitively on the inflaton potential.  Alternatively,
isocurvature quantum fluctuations in an
axion-like field  could produce
a quadrupole
anisotropy proportional to the gravitational
wave anisotropy, independent of the inflaton potential. These
could also produce large scale structure with more power on larger
scales than their adiabatic counterparts.

\baselineskip=21pt
\newpage

The observation of primordial fluctuations in the Cosmic
 Microwave Background by the DMR instrument aboard the
COBE \cite{smoot1} satellite provides one of the most important and
fundamental pieces of data in cosmology.  In principle,
the spectrum of fluctuations observed on large angular
scales can be extrapolated to smaller scales, where these
fluctuations would presumably have been responsible for
the formation of structure in the universe by
gravitational instability.  In normalizing this spectrum
to the observed COBE anisotropy however, one should be
aware of other possible contributions to this
anisotropy.  In particular, it was recently argued \cite{krauss}
that gravitational waves generated during an inflationary
phase in the early universe at plausible scales might
dominate the quadrupole anisotropy observed by COBE.
These waves contribute to the quadrupole and higher
moments of the CMB, while shorter wavelength modes
redshift away without affecting local observables, and
are thus not directly related to structure formation.
Thus, for example, small scale anisotropies (on
scales less than $1^o$)---which might directly probe
structure related density perturbations---and
corresponding large scale structure measurements may
be compared to the anisotropy observed by
COBE in trying to disentangle this situation
\footnote{During the preparation of this manuscript, I
learned from George Smoot about work of he and
collaborators \cite{smoot2} in which this issue and several others
I raise are treated carefully and in some detail. He
kindly made available to me their preliminary preprint
prior to submission.  While similarly motivated, I
believe this work is largely complementary to theirs. In
any case, I have tried to keep overlap to a minimum, and have
referenced accordingly.}.

There are
in fact independent reasons to suppose that a significant
component of the COBE anisotropy is related to
gravitational waves.  The standard CDM models of
structure formation, for example, generically predicts a
smaller quadrupole than the observations, by perhaps a
factor of at least O(2) \cite{bond,AbbWis2}.  It would be nice,
therefore, if there were a natural way in which scalar density
fluctuations and gravitational waves produced by
inflation could lead to comparable anisotropies on large
scales.\footnote{I thank Martin Rees for stressing the
interest in this possibility to me , and thereby
encouraging me to investigate it in more detail.}

The ratio of scalar to tensor modes predicted
to arise from inflation is model dependent.  The former
depends upon the detailed shape of the potential---in
particular on its first derivative, while the latter
depends merely on the energy density during the
inflationary epoch, and therefore on V(0), assuming this
 remains roughly
constant during inflation.   Utilizing standard analyses it is
straightforward to derive the ratio of the predicted quadrupole
anisotropies for various models (see also \cite{smoot2}).  Defining
in a conventional way the CMBR temperature anisotropy in terms of
spherical
 harmonics
\begin{equation}
{\delta T\over T}(\theta,\phi) =
\sum_{lm} a_{lm} Y_{lm}(\theta,\phi)
\label{eqn:mode-expansion}
\end{equation}
and projecting out a multipole to calculate the rotationally
symmetric quantity
\begin{equation}
  \left\langle a_l^2 \right\rangle \equiv
  \left\langle \sum_m |a_{lm}|^2 \right\rangle,
\end{equation}
the mean quadrupole anisotropy due to gravitational
waves from inflation can be simply given as \cite{Wisetc,krauss}
\begin{equation}
{\langle
a_2^2\rangle}\approx 7.7v.
\label{grav}
\end{equation}
where $v=V(0)/M_P^4$.  (The expectation value is based on a
statistical ensemble of universes which undergo
inflation, in one of which we happen to make a measurement.)
The predicted distribution of $a_2^2 $ is of a $\chi^2$ form with 5
degrees of freedom.

Alternatively, for exponential inflation one finds
that scalar density perturbations arising from the
inflaton field lead  to a quadrupole
anisotropy \cite{gys,Peebles,Linde1,Linde2}:
\begin{equation} {\langle
a_2^2\rangle}\approx {26.5 v^3 \over{v'}^2}.
\end{equation}
where $v'= {\left[{V'}/ M_P^3\right]^{2}}$ and $V'=dV/d\phi$ where
$\phi$ is the inflaton field.

The ratio, $R$, of these two quantities is then simply

\begin{equation} R\approx {0.29{v'}^2\over v^2} .
\end{equation}

It is clear that depending upon the nature of the
potential $R$ might vary substantially.
Indeed there exist a plethora of current inflationary
models including `new' inflation, chaotic inflation and extended
inflation.  Starobinsky
in fact calculated \cite{star} this ratio explicitly for
chaotic inflation models \cite{Linde2} and found $R\approx 0.25$
for a $\phi^4$ potential. This ratio has been concurrently re-examined in
detail for all inflationary models in \cite{smoot2}, where it is
pointed out that $R>1$ is possible but one is not completely free to
vary $v'$ independently of $v$.  For example, there exists a (weak) upper
bound on $R$ coming from the requirement that there be sufficient
inflation to solve the flatness and horizon problems.  Also, the
spectrum of density perturbations and CMB anisotropies will vary from an
n=1 Harrison-Zel'dovich spectrum when $R$ gets large.

Because existing inflationary models
can allow a wide variety of scalar to tensor
induced quadrupole anisotropies, observationally
distinguishing them could yield a
great deal of information about the mechanism of inflation(i.e.
see \cite{smoot2}).  However, for this same reason inflation does
not tend to generically predict them to be comparable in
magnitude. In addition, while inflation allows the
possibility that $R$ can be O(1) or greater, all existing
models suffer from the requirement that to produce
adiabatic density perturbations which are themselves sufficiently
small to be accomodated by the observed CMB anisotropies, various
parameters of the models already have to be tuned to be very
small. Whatever mechanism assures that they are small in the
first  place
might make them much smaller than the maximal value allowed by CMB
measurements.  Further tuning
of the potentials to obtain comparable tensor and scalar modes
does not make the situation any more palatable.

It should also be noted that because the contributions to the CMB
quadrupole anisotropy from adiabatic density perturbations and
gravitational waves are independent, they should be added in
quadrature to determine the final predicted result:

\begin{equation}
{\langle a_2^2\rangle}_{TOT}= {{\langle
a_2^2\rangle}_{scalar}} + {{\langle
a_2^2\rangle}_{tensor}}
\end{equation}
As soon as one term becomes slightly larger than the
other, it can quickly come to dominate the resulting
anisotropy.

Thus, while inflationary
potentials can produce comparable
anisotropies from adiabatic density perturbations and
gravitational waves, one wonders whether another
mechanism associated with inflation might also produce such a
situation.  I suggest one such possibility here,
which has the attractive feature that it may result in density
perturbations which may provide a better fit to large scale
structure observations than that resulting from adiabatic density
perturbations.

The suggestion is based on the calculation of the gravitational
wave anisotropy itself. Since the work of Grischuk \cite{grsh}, it
has been clear that gravitational wave generation during
inflation reduces to the calculation of the generation of
fluctuations for a massless scalar field in a background de Sitter
expansion. Each polarization state of gravitons behaves as a
massless, minimally coupled real scalar field, with a
normalization factor of $\sqrt{16\pi G}$ relating the two. As a
result, therefore, long wavelength fluctuations generated in any
massless (or light) scalar field will have an amplitude tied to
that for gravitational waves, up to a normalization
constant.

Quantum fluctuations in fields other than the
inflaton will result not in adiabatic density perturbations,
but rather in an isocurvature fluctuation
spectrum \cite{turner,linde}.  While the net effect on the
microwave background is remarkably similar (see below), the
impact for structure formation differs in important ways(i.e. see
\cite{turner,kolbturn}).  In particular, the final density perturbation
spectrum inside the horizon is flatter with wavenumber, with
relatively more power on larger scales than the adiabatic
spectrum.  This is also what recent observations
suggest \cite{efstath}.

Because isocurvature fluctuations involve no real energy density
perturbations (or their gauge invariant generalization)
 on scales larger than the horizon, fluctuations in
the number density of one species must be compensated by
fluctuations in the remaining species, including
radiation.  This results in fluctuations in the temperature of
the radiation \cite{kolbturn}.  If the scalar field $X$ comes to
dominate the energy density of the universe then the resulting induced
temperature fluctuations are identical to those induced had these
fluctuations been real energy density fluctuations.  Namely

\begin{equation}
{\delta T\over T} \approx {\kappa} {\delta \rho_X\over \rho_{tot}}
\approx \kappa \Omega_X {\delta \rho_X\over \rho_{X}}
\end{equation}
where $\delta \rho_X$ is the energy density density  in
the fluctuating field, $\rho_{tot}$ is the total energy
density and $\kappa \approx -1/3$ (X dominated) or $-1/4$
(radiation dominated).

For a massless field, the energy density contained in the
fluctuations is sufficiently small so that the effect on the
microwave background is minimal  (The energy density contained
in the gravitational waves generated by inflation, for example, is
small \cite{krauss}, but because of their direct
effect on the metric, they can produce an observable $\delta
T/T$.)  A simple estimate for the $ \delta T/T$ induced due to
horizon sized fluctuations in such fields today from
inflation at scale $v$ is then
\begin{equation}
{\delta T \over T} \approx {4 v \over 3}
\end{equation}

However, if the scalar field contributes significantly to the
energy density today, isocurvature fluctuations can contribute
both to structure formation, and to the CMB anisotropy. There are
two ways in which this might come about: (a) the scalar field was
massless during inflation, but received a non-zero mass at late
times.  A canonical axion is a prime example; or (b) the scalar
field had a non-zero, but small mass at all times.  This could
come about for a goldstone boson field with small explicit
symmetry breaking due to high energy effects at, say, the Planck
scale. We estimate the resulting quadrupole anisotropy below.

A de-Sitter expansion will produce for a massless, minimally coupled
scalar field a spectrum of fluctuations with Fourier components
 of
wavenumber $k$, which have crossed outside the horizon, given by
(i.e. \cite{BunchDav,gys,kolbturn,Linde2,krauss,white}):
\begin{equation}
{k^3|\delta \phi_k|^2\over2\pi^2} = \left( {H\over 2\pi}\right)^2
\nonumber
\end{equation}
where
\begin{equation}
\delta\phi_k\equiv \int d^3x\ \phi(\vec{x})
e^{i\vec{k}\cdot\vec{x}}
\end{equation}
These correspond to a background of $\phi$ particles with
occupation number \cite{Linde2}:
$n_k = {H^2/ 2k^2}$.

The energy density of the scalar field fluctuation is proportional
to $\phi^2$.  As a result, the fractional energy density stored
in the fluctuating field is ${\delta \rho(\phi) /
\rho(\bar{\phi})} =2 {\delta \phi /\bar{\phi}}$
and so the rms density fluctuations of wavenumber k, in
the $\phi$ field are given by
\begin{equation}
\left\langle \left({\delta \rho(\phi) \over
\rho(\bar{\phi})}\right)^2\right\rangle_k =4 {k^3 |\delta \phi_k|^2
\over 2\pi^2\bar{\phi}^2} ={H^2\over \pi^2 \bar{\phi}^2}
\end{equation}

This expression can then be inserted into standard formulas for
the induced quadrupole anisotropy \cite{AbbWis2,Peebles} to yield
\begin{equation}
\langle a_2^2 \rangle = ({160 \pi^2\over 9}) v \left[{M_P\over
\bar{\phi}}\right]^2
\label{result}
\end{equation}

We can now interpret this result in
terms of various possible particle physics models.  First,
comparing eq. (\ref{result}) to eq. (\ref{grav}), we see that the
two results are similar up to the overall
multiplicative factor $\left[{M_P/ \bar{\phi}}\right]^2$.  Thus,
if
\begin{equation}
\bar{\phi} \approx O(3)\  M_P,
\label{req}
\end{equation}
the scalar isocurvature
fluctuation induced quadrupole can be comparable to that induced
by gravitational waves.

Consider first a standard cosmic axion.  One finds that in this case the
condition (\ref{req}) cannot be easily satisfied.  This is
seen by re-writing the axion field in its canonical
angular form $ \bar{\phi} =\bar{\theta} f_a$, where $f_a$ is the
PQ symmetry breaking scale. First, in order for the axion to
exist at the inflationary scale, the PQ symmetry must break at or
above this scale.  For gravitational waves to contribute
significantly to the observed quadrupole the inflation scale must
be $>O(10^{16})$ GeV \cite{krauss}. It is well known that for $f_a
> 10^{12}$ GeV, $\bar{\theta}$ is required to be $\ll1$ in order
for axions not to overclose the universe.  Thus, for (\ref{req})
to be satisfied, apparently $f_a > M_P$ is required.  However,
even this is not sufficient, since in this case one would have to
fix $\bar{\theta} < \delta \theta$, which is not possible
\cite{turnwilcz}, nor would it be allowed by CMB constraints in any
case.

This problem for standard axions is intimately tied to the
validity of the standard relation between the axion mass and PQ
symmetry breaking scale.  This relation may not be correct.
It has been argued that it could changed
due to explicit global symmetry violation at the Planck
scale\cite{Georgi,Kamio}.  In this case (\ref{req})
might be enforceable.  More generally, other
pseudo-goldstone fields, related to spontaneous symmetry breaking
near the Planck Scale, with some (perhaps exponentially) small
explicit violation might play a role.  In either case, what is probably
required is a mass which is at least 3-4 orders of magnitude larger
than the canonical axion for a given scale of spontaneous symmetry
breaking.  One should also note that if such a field contributed only a
fraction of the closure density, this relation will be relaxed by the
same fraction.

Note that the masslessness of
a scalar field such as the axion at the PQ breaking scale is not
essential to the argument presented here.  It will apply equally
well to any scalar field, be it a pseudo-goldstone boson with a
 mass from small non-perturbative effects
arising at high energy--- perhaps the Planck Scale---
or any other ``light" scalar field.  As long as $m^2 \ll H^2$, the
above arguments go through essentially unchanged.  More
specifically, if $t < 3H/m^2$ \cite{Linde2}, where $t$ is the time
duration of the inflationary phase, all Fourier modes pushed outside the
horizon will have amplitudes identical to the massless case.  For
$V>10^{16}$ GeV this condition becomes, $ t_{max} \approx 2.5 \times
10^{-11} (GeV/m)^2 sec $, which is likely to be satisfied by any
inflationary phase at very early times, even for GeV
scale masses.  Of course, once the age of the universe becomes
comparable to $m^{-1}$, the magnitude of $|\delta \phi|^2$ will redshift
for all such modes, as the $\phi$ field begins to oscillate in its
potential, but so will any non-zero mean value of $\phi$.  Thus, the
fractional value of $\delta \phi /\phi$ or equivalently $\delta \rho
/\rho$ will not change.   Indeed this same phenomenon would presumably
apply to an axion background field.

Finally, some general comments are in order. As I have stressed here,
existing inflationary models predict three sources of CMB anisotropies:
adiabatic density fluctuations coming from quantum fluctuations in the
field driving inflation as it evolves in its potential, and
gravitational wave tensor perturbations and isocurvature density
perturbations coming from quantum fluctuations in spin 2 graviton and/or
spin 0 scalar fields present during inflation respectively.  The latter
two are determined by the vacuum energy density during inflation.  I have
concentrated primarily on the relationship between these two
components here. Nevertheless, one must recognize that in order for this
relationship to be pertinent, the CMB quadrupole anisotropy due to
adiabatic density fluctuations must be subdominant. Indeed, in the most
general case, the quadrupole anisotropy should be written:
\begin{equation} {\langle a_2^2\rangle}_{TOT}= {{\langle
a_2^2\rangle}_{adiabatic \  scalar}} + {{\langle
a_2^2\rangle}_{isocurvature \  scalar}} + {{\langle
a_2^2\rangle}_{tensor}}
\end{equation}

Thus it is not entirely
correct to state that the latter two terms can both be
significant and comparable, independent of the inflationary potential,
because in order to suppress the adiabatic component, as it
now stands one must inevitably tune the inflationary potential, with
the consequences for the shape of the power spectrum discussed in
\cite{smoot2}.  Nevertheless as I earlier stressed, in all existing
inflationary models there is an uncomfortable fine tuning required to
make adiabatic perturbations acceptably small.  It does not
seem unreasonable to suppose that in the
``correct'' model of inflation, if indeed there is such a thing, there
should be a mechanism which can naturally suppress, without fine
tuning, the density fluctuations induced as inflation ends and a
radiation dominated expansion begins---perhaps to values well below
the maximal allowed by CMB anisotropies.  Such a mechanism need not,
however, affect the latter two sources of anisotropy, which are not
directly sensitive to the transition from de Sitter to
radiation-dominated expansion.

These latter statements are more speculative.  However, until we
have a truly compelling inflationary model, which may
inevitably be tied to our understanding of Planck Scale
physics, we should not rule out such possibilities, and
perhaps should use them to get a handle on the actual
physical mechanism underlying inflation.  It is in this spirit
that the arguments I present here may be useful.

To conclude, inflation at present offers both the possibility of tying,
through a judiciously chosen potential, the magnitude of
anisotropies due to adiabatic density perturbations and
those due to gravitational waves, and {\it also} the
possibility that isocurvature perturbations in scalar fields
can arise which may also produce
comparable quadrupole anisotropies.  The latter two are tied together
independent of the potential, but depend sensitively upon the particle
physics parameters of the scalar fields.  If isocurvature fluctuations
are significant they could produce
a potentially interesting power spectrum for large scale structure.
It remains to see whether nature takes advantage of any of these
possibilities, or indeed of inflation itself.

\newpage
\noindent{I thank Martin Rees for sparking my interest
in this issue, and George Smoot and Martin White for helpful
discussions. George, Martin, and Paul Steinhardt also
offered useful comments and corrections to the original manuscript.
Note: I recently learned of a preprint by Dolgov and Silk \cite{silk},
which confirms the ratio\cite{star} of tensor and scalar induced CMB
anistropies in chaotic inflation in the context of COBE. }


\begin{thebibliography}{99}
\bibitem{smoot1}  G.F. Smoot {\it et al}, Astrophys. J. Lett., in press
(1992) \bibitem{krauss} L.M. Krauss and M. White, YCTP-P15-92, Phys.
Rev. Lett., in press.  The potential relation of gravitational waves
to possible COBE limits was also pointed out by G. F. Smoot in
NATO ASI Series
{\it Observational Tests of Cosmological Inflation} ed. by
Shanks et al.
 Kluwer Academic Publisher 1991
\bibitem{smoot2} R.L. Davis, H. M. Hodges, G. F. Smoot, P.J.
Steinhardt, M. S.Turner, FermiLab Pub-92/168-A, submitted PRL
\bibitem{bond} J.R. Bond and G. Efstathiou, Mon. Not. R. Astr. Soc.
{\bf 226} (1987) 655
\bibitem{AbbWis2} L.F. Abbott and M.B. Wise, Ap. J. {\bf 282} (1984) L47;
\bibitem{Wisetc} L.F.
Abbott and M.B. Wise, Nucl. Phys. {\bf B244} (1984) 541;
for earlier analyses see: V.A. Rubakov,
M.V. Sazhin and  A.V. Veryaskin,
Phys. Lett. {\bf B115} (1982) 189; A.A. Starobinsky, Sov. Astron.
Lett. {\bf 9} (1983) 302; A.A. Starobinsky, Sov. Astron. Lett.
{\bf 11} (1985) 133; R. Fabbri and M.D. Pollock,
Phys. Lett. {\bf B125} (1983) 445;
 R. Fabbri in Proceedings of the International School of Physics, Enrico Fermi,
1982, course 86, ed. F. Melchiorri and R. Ruffini (North-Holland): see
also L. P. Grishchuk, M. Solokhin, Phys. Rev. {\bf D43} (1991) 2566;
M. White, YCTP-P16-92, Phys. Rev. {\bf D}, in press
\bibitem{gys}  A. Guth and
S.-Y. Pi, Phys. Rev. Lett. {\bf 49} (1982) 1110; S. Hawking, Phys. Lett.
{\bf B115} (1982) 295; A.A. Starobinsky, Phys. Lett. {\bf B117}
(1982) 175; J. Bardeen, P. Steinhardt and M. Turner, Phys. Rev.
{\bf D28} (1983) 679; L.F. Abbott and M.B. Wise, Ap. J. {\bf 282}
(1984) L47
\bibitem{Peebles} P.J.E. Peebles, Ap. J. {\bf 263} (1982) L1
\bibitem{Linde1} A. D. Linde, Phys. Lett. {\bf B116} (1982) 335
\bibitem{Linde2} see A. D. Linde,{\it Particle Physics and Inflationary
Cosmology} (Harwood Pub., Chur 1982) and refs. therein
\bibitem{star} A.A. Starobinsky, Sov. Astr. Lett. {\bf 11} (1985) 133;
\bibitem{grsh} L.P. Grishchuk, Sov.Phys. JETP {\bf 40} (1975) 409;
L.P. Grishchuk, Ann. N.Y. Acad. Sci. {\bf 302} (1977) 439
\bibitem{turner} D. Seckel, M.S. Turner, Phys. Rev. {\bf D32} (1985)
3178
\bibitem{linde} A.D. Linde, Phys. Lett. {\bf B158} (1985) 375
\bibitem{kolbturn}  E.W. Kolb, M.S. Turner, {\it The Early Universe}
(Addison-Wesley, Redwood City, 1990)
\bibitem{efstath} S.J. Maddox {\it et al} Mon. Not. R. Astr. Soc.
{\bf 242} (1990) 43p
\bibitem{BunchDav} T.S. Bunch and P.C. W. Davies, Proc. R. Soc. London
{\bf A360} (1978) 117
\bibitem{white} M. White, YCTP-P16-92, Phys. Rev. {\bf D}, in press
\bibitem{turnwilcz}  M.S. Turner and F. Wilczek, Phys. Rev. Lett.
{\bf 66} (1991) 5
\bibitem{Georgi}  H. Georgi, L.J. Hall, M. B.Wise,
Nucl. Phys. {\bf 192} (1981) 409
\bibitem{Kamio} R. Holman {\it et al}, Phys. Let. {\bf B282} (1992) 132;
M. Kamionkowski and J. March-Russell, Phys. Let. {\bf B282} (1992) 137;
S.M. Barr and D. Seckel, BA-92-11 preprint (1992); see also R. Holman,
T.W. Kephart, S-J. Rey, YCTP -P17-92, for an explicit calculation, with
exponentially small violation.
\bibitem{silk} A. Dolgov and J. Silk, Berkeley preprint (1992)

\end{thebibliography}
\end{document}